\documentclass[12pt]{article}
\newcommand{\be}{\begin{equation}}
\newcommand{\ee}{ \end{equation}}
\newcommand{\ba}{\begin{array}}
\newcommand{\ea}{\end{array}}

\let\LARGE=\Large
\let\Large=\large

\def\unit{\hbox to 3.3pt{\hskip1.3pt \vrule height 7pt width .4pt \hskip.7pt
\vrule height 7.85pt width .4pt \kern-2.4pt
\hrulefill \kern-3pt
\raise 4pt\hbox{\char'40}}}

\begin{document}


\thispagestyle{empty}
\rightline{UUPHY/99/09}
\rightline{hep-th/9908048}
\vspace{1truecm}

\centerline{\bf \LARGE  Eguchi-Hanson metric from various limits}

\vspace{1truecm}

\centerline{{\bf Swapna Mahapatra}
\footnote{E-mail: swapna@iopb.res.in}} 

\vspace{.5truecm}

\centerline{\em Physics Department, Utkal University, 
Bhubaneswar-751004, India}


\vspace{.5truecm}

\begin{abstract}

\noindent 
In this note, we review various seemingly different ways of 
obtaining Eguchi-Hanson metric with or without a cosmological 
constant term. Interestingly, the conformal class of metric 
corresponding to hyperbolic $n$-monopole solution obtained 
from the generalized Gibbons-Hawking ansatz, reduces to the 
Eguchi-Hanson metric in a particular limit. These results, 
though known from an algebraic geometry point of view,
are useful while dealing with rotational killing symmetry 
of self-dual metrics in general theory of relativity as well 
as in the context of duality symmetry in string theory.

\end{abstract}


\newpage


Gravitational instanton solutions in the context of Euclidean 
gravity and string theory have become the subject of much 
interest in recent times. They are defined to be nonsingular, 
complete, positive definite (Riemanninan metric) solutions of 
vacuum Einstein equations or Einstein equations with a 
cosmological constant term \cite{Eguchi1, Gibbons1}. The 
existence of 
such solutions is important in the study of quantum theory
of gravity. These are analogous to Yang-Mills 
instantons \cite{Belavin}, which are defined as nonsingular 
solutions of classical equations in four dimensional Euclidean 
space. The Yang-Mills instantons 
are characterized by self-dual field strengths, whereas the 
gravitational instantons are characterized by self-dual or 
anti self-dual curvature.
There are also examples of gravitational instantons which are 
not self-dual, those are the Euclidean version of Schwarzschild 
and Kerr solutions. The four dimensional Riemannian manifolds 
(${\cal M}, g_{ab}$) for gravitational instantons can be 
asymptotically flat (AF), asymptotically
locally Euclidean (ALE), asymptotically locally flat (ALF) or 
compact without boundary. Multi Taub-NUT solution of Hawking 
\cite{Hawking2}
is an example of ALF space. The simplest nontrivial example of ALE 
spaces is the metric of Eguchi-Hanson \cite{Eguchi2}.  
ALE instantons have been found explicitly by Gibbons and 
Hawking \cite{Gibbons2} and they are known implicitly through
the work of Hitchin \cite{Hitchin1} 
The complex projective space
$CP^2$ is an example of compact, anti self-dual instanton solving 
Einstein's equation with a cosmological constant term 
\cite{Cp}. 
 
There are two topological invariants associated with these 
solutions, namely the Euler characteristic
$\chi$ and the Hirzebruch signature $\tau$, which can be 
expressed as integrals of the curvature of a four dimensional 
metric. The topological invariants are also related to nuts 
(isolated points) and bolts (two 
surfaces), which are the fixed points of the action of one 
parameter isometry groups of gravitational instantons.
The Eguchi-Hanson metric is given by,

\begin{eqnarray}
d s^2 & = & {(1 - \frac{a^4}{r^4})}^{-1} d r^2 + 
r^2(\sigma_x^2 + \sigma_y^2)
+ r^2 (1 - \frac{a^4}{r^4})\sigma_z^2
\end{eqnarray}
In terms of Euler angles $\theta, \phi$ and $\psi$, the differential
one forms $\sigma_i$ are expressed as,
\begin{eqnarray}
2 \sigma_x &=& (\sin\psi d\theta - \sin\theta \cos\psi d\phi)
\\
2 \sigma_y &=& (- \cos\psi d \theta - \sin\theta \sin\psi d\phi)
\\
2 \sigma_z &=& (d\psi + \cos\theta d\phi)
\end{eqnarray}
This metric has a single removable bolt singularity provided $\psi$ lies 
in the range $0 < \psi < 2\pi$. Asymptotically the topology of the 
manifold is $S^3/Z_2$ which is not globally Euclidean. Near $r=a$,
the manifold has the topology $R^2 \times S^2$. 
The self dual Euclidean Taub-NUT solution
is given by,
\begin{eqnarray}
d s^2 &=& \frac{1}{4} (\frac{r + m} {r - m}) d r^2 + 
\frac{1}{4} (r^2 - m^2) (\sigma_x^2 + \sigma_y^2) + 
m^2 (\frac{r - m}{r + m})\sigma_z^2
\end{eqnarray}
This metric has a single removable nut singularity. Both 
the above metrics 
have self-dual Riemann curvature, where the dual of the 
Riemann tensor $R_{i j k m}$ is defined as,
\begin{eqnarray}
\star R_{i j k m} &=& \frac{1}{2}{\sqrt g}
\epsilon_{k m r s} R_{i j}^{r s}
\end{eqnarray}
The self(anti)-duality condition is given by,
\begin{eqnarray}
\star R_{i j k m} &=& \pm R_{i j k m}
\end{eqnarray}
\\
The Fubini-Study metric on $CP^2$ is given by,
\begin{eqnarray}
d s^2 &=& \frac{d r^2 + r^2 \sigma_z^2}{(1 + \frac{\Lambda r^2}{6})^2}  
+ \frac{r^2(\sigma_x^2 + \sigma_y^2)}{(1 + \frac{\Lambda r^2}{6})}
\end{eqnarray}
where, $\Lambda$ is the cosmological constant. This metric 
has an anti
self-dual Weyl tensor. The metric has a nut as well as a 
bolt type of sungularity. All these three metrics can be derived 
from a more general three parameter Euclidean 
Taub-NUT de Sitter metric through some limiting procedure. 
This limiting procedure to obtain the corresponding solutions in 
string theory has been discussed in \cite{Swapna}.
\\
The Taub-NUT de sitter metric is given by,
\begin{eqnarray}
d s^2 &=& \frac{\rho^2 - L^2}{4 \Delta} d\rho^2 + (\rho^2 - L^2)
(\sigma_x^2 + \sigma_y^2) + \frac{4 L^2 \Delta}
{\rho^2 - L^2}\sigma_z^2
\end{eqnarray}
where,
$\Delta = \rho^2 - 2 M\rho + L^2 + \frac{\Lambda}{4}
(L^4 + 2 L^2 \rho^2 - \frac{1}{3}\rho^4)$.
One limit which has been discussed in \cite{Eguchi3}
is to set 
\begin{eqnarray}
M &=& L(1 + \frac{a^4}{8 L^4} + \frac{\Lambda L^2}{3})
\end{eqnarray}
in the above. Putting $\Lambda = 0$ and then taking the limit 
$L \rightarrow\infty$ with $r^2 = \rho^2 - L^2$ held fixed, 
one obtains the Eguchi-
Hanson metric. If one does not put $\Lambda = 0$, then it 
reduces to the Eguchi-Hanson de Sitter metric given by,
\begin{eqnarray}
d s^2 &=& {(1 - \frac{a^4}{r^4} - \frac{\Lambda r^2}{6})}^{-1}
d r^2 + r^2 (\sigma_x^2 + \sigma_y^2) + r^2 (1 - \frac{a^4}{r^4} - 
\frac{\Lambda r^2}{6}) \sigma_z^2
\end{eqnarray}
This metric satisfies Einstein equation with a positive 
cosmological 
constant. An analogous solution has been derived by Pedersen 
\cite{Pedersen1} and the solution has been interpreted as a 
nonlinear 
superposition of Eguchi-Hanson and pseudo Fubini-Study 
solution. There is an apparent singularity where 
$1 - \frac{a^4}{r^4} - \frac{\Lambda r^2}{6} = 0$  
which can be removed by adjusting the parameter $a$.

The above solution can also be derived as a  
generalization of the vacuum Einstein solution obtained 
through 
a twistor theoretical method \cite{Pedersen2}. The vacuum 
solution was of the form,
\begin{eqnarray}
d s^2 &=& \frac{1}{4 R^3} [(d R - 2 m R^2 \sigma_z)^2 + 4 R^2 
(\sigma_x^2 + \sigma_y^2 + \sigma_z^2)]
\end{eqnarray}
This vacuum solution has been obtained by generalizing 
Lebrun's work on H-spaces with cosmological 
constant \cite{Lebrun1} to the Berger sphere (3-sphere 
squashed in one 
direction). The metric on the Berger sphere is given by, 
\begin{eqnarray} 
d s^2 &=& d \theta^2 + \sin^2\theta d\phi^2 + 
\cos^2\alpha (d\psi + \cos\theta d\phi)^2
\end{eqnarray}
where $\alpha$ is a fixed constant. For $\alpha = 0$, 
the metric reduces to that of the metric
on 3-sphere $S^3$. More generally, the metric on 
Berger sphere can be written as $\sigma_x^2 + \sigma_y^2 + 
I_3 \sigma_z^2$, where $I_3$ is a constant. The null 
geodesics of 
the Berger sphere also describe the motion of a symmetric 
top where
$I_3$ can be recognized as the moment of inertia along the 
third body axes. The complex three dimensional
twistor space $z$ of unparametrized null geodesics of the 
Berger sphere is described by a trivial line 
bundle over plane sections of the quadric.  
The line bundle then generates a 
conformal structure and a $U(1)$ monopole on $S^3$ 
characterized
by the Higgs field $V$ and a gauge potential $A$. The 
conformal structure is given by, 
\begin{eqnarray}
g &=& V(d\chi^2 + \sin^2\chi(d\theta^2 + \sin^2\theta d\phi^2)) +
V^{-1}(d\tau + A)^2
\end{eqnarray}

Here the monopole solution is characterized by $(V, A) = 
(\epsilon + m \cot\chi, m\cos\theta d\phi)$ and one takes 
$\epsilon = m^2 = I_3^{-1} -1$. We do not go 
to the discussion 
on the construction of the twistor space of unparametrized 
null geodesics of the Berger sphere.  
For details on the construction of the twistor space  
see \cite{Pedersen2}. Here we are interested in the 
corresponding solution of the 
Einstein equation with a cosmological constant which is given by,
\begin{equation}
g = F(\chi)^2 [(\epsilon + m\cot\chi) (d\chi^2 + 
\sin^2\chi(d\theta^2 + \sin^2\theta d\phi^2))   
\end{equation}
\[
+ \frac{m^2}{\epsilon + m\cot\chi} (d\psi + \cos\theta d\phi)^2] 
\]
where $\epsilon$ and $m$ are two parameters which can be adjusted
appropriately. Solving the Einstein's equations with a cosmological
constant term $(\sum_l R_{l i l j} = \Lambda \delta_{i j})$, 
one obtains,
\begin{eqnarray}
F(\chi) &=& \frac{k}{(\epsilon\cos\chi - m\sin\chi)^2}
\end{eqnarray}
where $k = -\frac{3\epsilon}{\Lambda}$. This solution has a self-dual 
Weyl tensor.
By making suitable coordinate transformation, the metric can be
simplified to a desired form,
\begin{equation}
g = \frac{k}{(\epsilon - m^2 R)^2} \big [\frac{m^2(1 + \epsilon R)}
{(1 + m^2 R^2)R} d R^2 + 
4 m^2 R(1 + \epsilon R)(\sigma_x^2 + \sigma_y^2) 
\end{equation}
\[
+ \frac{4 m^2 R(1 + m^2 R^2)}{1 + \epsilon R} \sigma_z^2 \big ]
\]
For $k = m^2/4$, $\epsilon = m^2$ and substituting $R = \rho^2$, 
one obtains the self-dual metric (which is conformally equivalent to 
the metric on the Berger sphere) as \cite{Pedersen2},
\begin{eqnarray}
\frac{1}{(1 - \rho^2)}\left [\frac{1 + m^2\rho^2}
{1 + m^2\rho^4}d \rho^2 
+ \rho^2 (1 + m^2 \rho^2)(\sigma_x^2 + \sigma_y^2) + 
\frac{\rho^2 (1 + m^2 \rho^4)}{1 + m^2 \rho^2}\sigma_z^2 \right ]
\end{eqnarray}
Conformally on $\rho = 1$ (defines the berger sphere), $g$ is 
given by, 
\begin{eqnarray}
(1 + m^2) (\sigma_x^2 + \sigma_y^2 + I_3 \sigma_z^2)  
\end{eqnarray}

For $\epsilon = 0$ and $k = m^2/4$, the cosmological constant 
goes to zero and the new metric reduces to that of Eguchi-Hanson I 
(identifying 
$m^2$ with $a^4$). Though the above derivations assume $I_3 < 1$ 
and $m^2 > 0$, one can relax this constraint and get a valid solution
for $m^2$ negative.    
Taking $m^2 = - a^4$ and $R = r^{-2}$, the vacuum solution
is of the form,
\begin{eqnarray}
d s^2 &=& (d r - \frac{i a^2}{r}\sigma_z)^2 + r^2(\sigma_x^2
+ \sigma_y^2 + \sigma_z^2)
\end{eqnarray}
A more general solution of Einstein's equation with a 
cosmological constant can be written
as \cite{Pedersen1}
\begin{eqnarray}
d s^2 &=& [d r - i g(r)\sigma_z]^2 + r^2(\sigma_x^2 + 
\sigma_y^2 + \sigma_z^2)
\end{eqnarray}
By solving the Einstein equation, one can show that the 
solution is given by 
$g(r) = (a^4/r^2 + b^2 r^4)^{1/2}$ and it satisfies the 
Einstein's equation 
with $\Lambda = 6 b^2$. When $b= 0$, the above metric 
reduces to the Eguchi-Hanson solution. This can be seen by 
defining a new function 
$d H(r) = \frac{2 i a^2}{r}(r^2 - \frac{a^4}{r^2})^{-1} d r$ 
and making a coordinate transformation $\hat\psi
= \psi - H(r)$. This brings the metric to the standard form,
\begin{eqnarray}
d s^2 &=& (1 - \frac{a^4}{r^4})^{-1} d r^2 + r^2(\rho_1^2 + 
\rho_2^2) + r^2 (1 - \frac{a^4}{r^4})\rho_3^2
\end{eqnarray}
where, 
\begin{eqnarray}
2\rho_3 &=& d\hat\psi + \cos\theta d\phi 
\\
4(\rho_1^2 + \rho_2^2) &=& (d\theta^2 + \sin^2\theta d\phi^2)
\end{eqnarray}
Similarly, in the limit $a = 0$, the metric reduces to,
\begin{eqnarray}
d s^2 &=& (d r - i b r^2 \sigma_z)^2 + r^2 (\sigma_x^2 + \sigma_y^2
+ \sigma_z^2) 
\end{eqnarray}
Again defining a new function 
$d K(r) = 2 i b (1 - b^2 r^2)^{-1} d r$
and putting $\hat\psi = \psi - K(r)$, one obtains the 
(pseudo) Fubini-Study metric on $CP^2$ given by, 
\begin{eqnarray}
d s^2 &=& \frac{d R^2 + R^2 \sigma_z^2}{(1 - 
\frac{\Lambda R^2}{6})^2} 
+ \frac{R^2(\sigma_x^2 + \sigma_y^2)}{1 - 
\frac{\Lambda R^2}{6}}, \qquad \qquad \Lambda > 0
\end{eqnarray}
where we have substituted $\Lambda = 6 b^2$ and the 
coordinate $R$ 
has been defined as $r^2 = R^2(1 - \frac{\Lambda R^2}{6})^{-1}$.
For both $a$ and $b$ nonzero, one can again define a new function and 
make a coordinate transformation such as,
\begin{eqnarray}
d G(r) &=& \frac{2 i}{r}(\frac{a^4}{r^4} - b^2 r^2)^{1/2} (1 - \frac
{a^4}{r^4} - b^2 r^2)^{-1} dr, \qquad \qquad \hat\psi = \psi - G(r)
\end{eqnarray}
and the above metric is then recognized as the Eguchi-Hanson de 
Sitter metric given by,
\begin{eqnarray}
d s^2 &=& (1 - \frac{a^4}{r^4} - \frac{\Lambda r^2}{6})^{-1} dr^2 
+ r^2 (\rho_1^2 + \rho_2^2) + r^2(1 - \frac{a^4}{r^4} - 
\frac{\Lambda r^2}{6})\rho_3^2, \qquad \Lambda > 0
\end{eqnarray}
This is precisely the metric one obtains from Taub-NUT de Sitter
solution through a singular limiting procedure as we 
discussed before. 

The above metric has an apparant singularity where $\Delta = 0$,
which can be removed by adjusting the parameter $a$. Here, for 
example, the removable bolt singularity occurs at 
\begin{eqnarray}
r &=& {\sqrt\frac{2(n - 2)}{\Lambda}} , \qquad \qquad n \geq 3
\end{eqnarray}
This metric
has the interpretation of the nonlinear superposition of the 
Eguchi-Hanson metric II and the pseudo Fubini-Study metric. 
Note that the metric of the form 
\begin{eqnarray}
d s^2 &=& [d r - (\frac{a^4}{r^2} + b^2 r^4)^{1/2} \sigma_z]^2
+ r^2(\sigma_x^2 + \sigma_y^2 + \sigma_z^2) 
\end{eqnarray}
satisfies Einstein's equation with a negative cosmological 
constant ($\Lambda = - 6 b^2$), which again reduces to 
Eguchi-Hanson 
for $b=0$ and (pseudo) Fubini-Study solution for 
$a = 0$ \cite{Pedersen1}.

Next we discuss a very different method of obtaining the 
ALE metric
as a limit of the self-dual metrics on the connected sum 
of $nCP^2$
(which is related to the hyperbolic $n$-monopole solution). 
Gibbons and Hawking have given
the ansatz for the multi-center metric in 
the form
\begin{eqnarray}
d s^2 &=& V^{-1}({\bf x})(d\psi + \omega \cdot d{\bf x})^2 + 
V({\bf x}) \gamma_{i j} d{\bf x}^i \cdot d{\bf x}^j 
\end{eqnarray}
with $\nabla V = \pm \nabla \times \omega$. $V$ satisfies 
the three
dimensional Laplace equation $\nabla^2 V = 0$, whose 
most general solution is given by,
\begin{eqnarray}
V({\bf x}) = \epsilon + \sum_{i = 1}^n \frac{m_i}
{|{\bf x} - {\bf x}_i|}
\end{eqnarray}
where $\epsilon$ and $m_i$ are constants. For removable 
nut singularities,
one takes all the $m_i$ to be equal ($m_i = M$) and $\psi$ 
sould be periodic 
with the range $0 \leq \psi \leq 8\pi M/n$. The value  
$\epsilon = 1$
corresponds to ALF metrics (the self-dual Taub-NUT solution 
corresponds to $n=1$). For $\epsilon = 0$, one obtains the ALE 
metrics ($n=1$ corresponds to flat space, $n=2$ corresponds to 
Eguchi-Hanson and so on). The ALE and ALF metrics are related 
to each other through a sequence of $T-S-T$ duality transformation 
(more precisely through Ehler's transformation) \cite{Bakas1}.
 
There has been attempt to find all 
the Riemannian metrics with atleast one killing vector field
\cite{Das}. There are two classes of these solutions 
depending on the type of killing symmetry. One class of metrics
admits translational killing vectors and is determined by $V$ 
satisfying the 3-dimensional Laplace equation in Euclidean space.
The other class of metrics admit rotational killing symmetry, 
where the metric is determined by a scalar field satifying a non
linear partial differential equation, namely the continual 
Toda equation. In the above form of the metric, $\omega_i$ are
the components of connection 1-form on the three dimesional
Riemannian manifold. The distinction between the translational 
and rotatonal killing symmetry is made by examining whether the 
quantity $\Delta\psi \equiv \gamma_{i j}\partial_i\psi\partial_j
\psi = 0$ (translational) or $\Delta\psi > 0$ (rotational). Here 
$\psi$ is a scalar field consisting of the nut potential and the 
$V$ appearing in Gibbons-Hawking ansatz. In string theory context, 
this scalar field is a combination of the axion and dilaton fields
($ S_{\pm} = b \pm e^{-2\Phi}$). Choosing one of these 
conjugate fields
$S_-$, the constraint is whether $\Delta S_- = 0$ or $\Delta 
S_- > 0$.  
This analysis is useful while dealing with the sequence of 
$T-S-T$ duality transformations. This sequence of duality
transformations take a pure gravitaional background to another 
pure gravitational background with a Ricci flat metric 
\cite{Bakas1}. In the 
rotational killing symmetry case, the self duality of the original
background gets broken, wheras the self-duality condition is 
preserved if one performs the T-duality transformation with respect
to the translational killing symmetry. A supersymmetric explanation
for the violation of self-duality has been provided in 
\cite{Bakas2}.

In general theory of relativity,
the previously discussed nonlinear differential equations 
are the heavenly eqations
which arise in the context of self-dual Einstein spaces with one 
rotational killing vector. Interestingly, these equations also 
arise in the context of problems in differential geometry, namely
the construction of half-flat metric on the connected sum of $n$
copies of the complex projective space $CP^2$. For compact, 
self-dual manifolds $X$ with positive scalar curvature, a 
theorem in 
Riemannian geometry says that $X$ is isometric to the Euclidean 
4-sphere or the complex projective plane $CP^2$ with the 
Fubini-Study metric. There has been lot of work to enlarge 
this set of 
examples of compact, self-dual manifolds (here self-duality 
means the anti-self-dual part of the Weyl curvature $W^-$ 
vanishes) apart from the conformally flat $S^4$, $CP^2$ and $K_3$. 
For example, using 
twistor methods, Poon has constructed families of self-dual 
structures on the connected sum of $2 CP^2$ ($CP^2 \# CP^2$)
and $3 CP^2$ ($CP^2 \# CP^2 \# CP^2$)
\cite{Poon}. Donaldson and Friedman have given 
general conditions under which the connected sum of two self-dual
Riemannian 4-manifolds again admit a self-dual structure. They have
also given a twistor technique for constructing self-dual structure
on the connected sums \cite{Donaldson}. 
If $X_1$ and $X_2$ are two self-dual manifolds with twistor spaces
$Z_1$ and $Z_2$ (3-dimensional complex manifolds) respectively, 
then the idea is to look for metrics in the connected sum 
$X_1 \# X_2$ which is close to the original metric outside a 
small neck
where the connected sum is made. Using the gluing procedure, 
Floer has also shown the existence of conformal structures with 
self-dual Weyl tensor in the connected sum of $n$ copies of 
complex projective space $CP^2$, ($n > 0$) \cite{Floer}. 
The conformal 
structures can be represented by metrics of positive scalar 
curvature. Here we should mention that a conformal structure 
on a smooth finite dimensional manifold ${\cal M}$ is an 
equivalence class $[g]$ of Riemannian metrics $g$ on ${\cal M}$,
where $g_1$ and $g_2$ are conformally  equivalent if 
there exists a non vanishing function $f$ such that 
$g_2 = f g_1$. 
In four dimensions, the Weyl
tensor $W$ can be decomposed in an invariant way to $W_+ + W_-$.
The conformal structure $[g]$ is self-dual if $W_- = 0$ and hence
it satisfies half of the integrability condition. In the twistor 
approach, one constructs a certain complex space $Z$ using $Z_1$
and $Z_2$ and looks for twistor spaces made by small smoothings
of $Z$. If such smoothings exist, then they always represent the
twistor spaces of self-dual structures on the connected sum. 

Explicit self-dual metrics on the connected 
sum of $nCP^2$ has been constructed by Lebrun \cite{Lebrun2}. 
The construction generalizes 
the Gibbons-Hawking ansatz for obtaining Ricci flat hyperk\"ahler
manifolds. Basically the three dimensional Euclidean space gets 
replaced by the hyperbolic 3-space and metrics are conformally 
K\"ahler with vanishning scalar curvature. The extra ingredient 
here is the circle ($S^1$) isometry. Let us elaborate a little 
bit on this. The self-dual Yang-Mills equation on $R^4$ 
(giving instanton solutions) and the Bogomolnyi equation 
on $R^3$ (giving monopole solutions) are respectively given 
by,
\begin{eqnarray}
\star F &=& F \\
D V &=& \star F 
\end{eqnarray}
where $V$ is the Higgs field and $F$ is the curvature of a 
connection $A$, $D$ is the covariant derivative {\it w.r.t.}
$A$ and $\star$ is the duality operator with respect to the 
Euclidean metric. One knows that the solutions of self-dual 
Yang-Mills equations which are independent of the coordinate 
$x_4$ (translationally invariant) reduce to the solution of 
the Bogomolnyi equation in $R^3$. If instead, we consider 
rotationally invariant solutions of the self-dual Yang-Mills 
equations in $R^4$, then the Euclidean metric can be rewritten 
as, 
\begin{eqnarray}
d s^2 &=& r^2 \left [ \frac{d x_1^2 + d x_2^2 + d r^2}{r^2} 
+ d \theta^2 \right ] 
\end{eqnarray}
where we have used polar coordinates $r$ and $\theta$ for 
the rotational invariance relative to angular rotation in 
the $(x_3, x_4)$ plane. The first factor on ${\it r.h.s}$
corresponds to the metric on the hyperbolic 3-space $H^3$
(can be represented as the upper-half space $z > 0$ in 
$x$, $y$, $z$ coordinates) of 
constant curvature $-1$. The above equation implies a 
conformal
equivalence (with $r^2$ as the conformal factor),
\begin{equation}
R^4 - R^2 \simeq H^3 \times S^1
\end{equation}
This means that $S^1$ invariant solutions (solutions 
independent
of $\theta$) of the self-dual Yang-Mills equations reduce
to the magnetic monopole solutions (solutions of Bogomolnyi
equation) on the hyperbolic 3-space $H^3$ \cite{Atiyah}. 
This conformal 
equivalence is a special case of more general equivalence 
$R^n - R^m \simeq H^{m + 1} \times S^{n - m - 1}$, $m < n$.  

The generalized Gibbons-Hawking ansatz is given by \cite
{Lebrun2},
\begin{eqnarray}
g &=& e^u w (d x^2 + d y^2) + w d z^2 + w^{-1} f^2
\end{eqnarray}
where $g$ is a scalar flat K\"ahler metric. $f$ is the 
connection 
1-form, $u$ and $w$ are two smooth real valued functions on 
an open set $R^3$ and they satisfy the partial differential
equations,
\begin{eqnarray}
u_{x x} + u_{y y} + {(e^u)}_{z z} &=& 0
\\
w_{x x} + w_{y y} + {(w e^u)}_{z z} &=& 0
\end{eqnarray}
One can recognize these two equations as the continual Toda 
equations
and its linearlizations. Note that  
the generalized  ansatz reduces to that of the 
Gibbons-Hawking ansatz if we set $u=0$. Then the above 
equations just reduce to the
3-dimensional Laplace equation as discussed before. 

In the context of the generalized ansatz, 
consider the K\"ahler form on $C^2 - \{\vec 0\}$ which 
is given by,
\begin{eqnarray}
\Omega &=& - \frac{i}{2} \partial \bar\partial 
(||z||^2 + m \log ||z||^2)
\end{eqnarray}
The K\"ahler potential is given by,
\begin{eqnarray}
\phi &=& \frac{1}{2}(||z||^2 + m \log ||z||^2)
\end{eqnarray}
where, ($||z||^2 = z_1 \bar z_1 + z_2 \bar z_2$) and $m$ is 
a positive constant. 
The above K\"ahler form defines a zero scalar curvature 
K\"ahler metric. 
This follows from the theorem of Lebrun \cite{Lebrun3} that 
for 
${\cal M} = m CP^2$, $0\leq m \leq 3$ equipped with a 
self-dual metric $g$ of positive scalar curvature, there 
exists atleast one point $p \in {\cal M}$ such that 
$({\cal M} - \{p\}, g)$ is conformally isometric to $C^2$ 
with $m$ points blown up equipped with an asymptotically flat 
K\"ahler metric of zero scalar curvature. The above K\"ahler 
form on $C^2 - \{0\}$ corresponds to the case $m = 1$. 
Computation of the volume form of the above    
metric gives 
$u = \log 2 z$, which is invariant under translations in 
both $x$ and $y$. Using this, the second equation reduces to,
\begin{eqnarray}
w_{x x} + w_{y y} + {(2 z w)}_{z z} &=& 0
\end{eqnarray}
Defining a new variable $V = 2 z w$, this equation just
reduces to the Laplace-Beltrami equation $\nabla V = 0$,
where, $\nabla$ is the Laplace-Beltrami operator of the 
Riemannian metric in the upper half space. The metric 
on the upper half space $\{(x, y, z) \in R^3 |z > 0 \}$ 
is given by,
\begin{eqnarray}
h &=& \frac{d x^2 + d y^2}{2 z} + \frac{d z^2}{4 z^2}
\end{eqnarray}
In terms of another coordinate $q = {\sqrt 2z}$, the 
above metric becomes,
\begin{eqnarray}
h &=& \frac{d x^2 + d y^2 + d q^2}{q^2}
\end{eqnarray} 
which is infact the metric on the hyperbolic 3-space $H^3$. 
Now using $h$ and the new variable $V$, the generalized 
Gibbons-Hawking ansatz reads,
\begin{eqnarray}
g &=& 2 z(V h + V^{-1} f^2)
\end{eqnarray}

One can show that in terms of the hyperbolic distance 
$\rho$ from the 
point $(x, y, q) = (0, 0, q_0)$ in the upper half space, $V$ 
corresponds to a single hyperbolic monopole solution,
\begin{eqnarray}
V = 2 z w = 1 + \frac{1}{2}({\tanh^{-1}\rho} - 1) = 
1 + \frac{1}{e^{2\rho} - 1}
\end{eqnarray}
where,
\begin{eqnarray}
\rho &=& \cosh^{-1}\left [\frac{x^2 + y^2 + q^2 + q_0^2}
{2 q q_0} \right ]
\end{eqnarray}
Also,
\begin{eqnarray}
w^{-1} &=& |z_1|^2 \left ( 1 + \frac{m |z_2|^2}{{(|z_1|^2 + 
|z_2|^2)}} \right ) \\
e^u &=& |z_1|^2 \left ( 1 + 
\frac{m}{|z_1|^2 + |z_2|^2} \right ) \\
|z_1|^2 &=& \frac{1}{2} \left [ q^2 - m - x^2 - y^2 
+ {\sqrt{(q^2 - m + x^2 + y^2)^2 + 4 m (x^2 + y^2)}} \right ] \\
|z_2|^2 &=& x^2 + y^2 
\end{eqnarray}
Considering an arbitrary collection of points $p_j$ 
in the upper half space $H^3$, with 
\begin{eqnarray}
V &=& 1 + \sum_{j = 1}^{n}\frac{1}{e^{2\rho_j} - 1}
\end{eqnarray}

one can show that the metric $g = q^2 [V h + V^{-1}  f^2]$ 
represents a scalar flat, asymptotically flat K\"ahler metric
on the blow up of $C^2$ at $n$-points \cite{Lebrun2}. 
The conformal 
class of the above metric represents the self-dual metric 
on the connected sum of $n CP^2$, whose scalar curvature 
is positive. There exists an interesting limit of this 
conformal metric.
If all the centres $p_1, \ldots p_n$ coincide to a single 
point, it defines a self-dual compact orbifold with an 
isolated singular point $p$. Interestingly, the space 
$H^3 - {p}$ with a suitable 
conformal factor gives rise to self-dual ALE manifold, 
which we discuss below.

Identifying the space $H^3 - {p}$ with $R^{+} \times S^2$, 
the hyperbolic metric can be written as,
\begin{eqnarray}
h &=& d \rho^2 + \sinh^2\rho g_{S^2}
\end{eqnarray}
where $g_{S^2}$ is the metric on the 2-sphere and $\rho$ is 
the radial coordinate. Since all the centres coincide, 
$V = 1 + \frac{n}{e^{2\rho} - 1}$. The circle bundle can be 
identified with 
$R^+ \times (S^3/Z_n) \rightarrow R^+ \times S^2$. 
By considering the projection from the lens space 
$S^3/Z_n$ to $S^2$ induced by the Hopf map, the connection 
1-form can be suitably chosen to be $f = - n \sigma_z$ 
\cite{Lebrun2}.
Substituting the expressions for $V$ and $h$, one finds that
the metric corresponding to generalized Gibbons-Hawking 
ansatz is conformally equivalent to (the conformal factor
being $4\sinh^2\rho$),
\begin{eqnarray}
g &=& \frac{e^{2\rho}(e^{2\rho} + n - 1)}{(e^{2\rho} - 1)^3} d\rho^2
+ \frac{e^{2\rho} + n -1}{e^{2\rho} - 1}(\sigma_1^2 + \sigma_2^2) +
\frac{n^2 e^{2\rho}}{(e^{2\rho} - 1)(e^{2\rho} + n - 1)}\sigma_3^2
\end{eqnarray}
Putting, $r = {\sqrt\frac{e^{2\rho} + n - 1}{e^{2\rho} - 1}}$,
one obtains,
\begin{eqnarray}
g &=& {(1 + \frac{A}{r^2} + \frac{B}{r^4})}^{-1} d r^2 + 
r^2 \left [\sigma_x^2 + \sigma_y^2 + (1 + \frac{A}{r^2} + 
\frac{B}{r^4})\sigma_z^2 \right ]
\end{eqnarray}
where, $A = n -2$ and $B = 1 - n$.
One can see that for $n = 2$, we have $A = 0$ and $B = -1$ and 
the metric precisely reduces to the Eguchi-Hanson metric in 
the standard form with the integration constant $a = 1$:
\begin{eqnarray}
g &=& (1 - \frac{1}{r^4})^{-1} d r^2 + r^2 (\sigma_1^2 + 
\sigma_2^2) + r^2(1 - \frac{1}{r^4})\sigma_3^2
\end{eqnarray}
This limiting procedure can be understood from the
hyperbolic space structure here. The hyperbolic metric of 
constant negative curvature automatically arises by 
considering 
rotational killing symmetry. By varying the curvature 
of the 
hyperbolic space and letting it tend to zero, the Euclidean 
space appears naturally as a limit of the hyperbolic 3-space.
We can rewrite the metric on $H^3 \times S^1$ as,
\begin{eqnarray}
d s^2 &=& \frac{r^2}{R^2}\left [\frac{R^2(d x^2 + d y^2 + 
d r^2)}{r^2}
+ R^2 d\theta^2 \right ]
\end{eqnarray}
The conformal equivalence in this case becomes $R^4 - R^2 
\sim H^3(R) \times S^1(R)$, where $S^1(R)$ is the circle of 
radius $R$ and $H^3(R)$ is the hyperbolic space of curvature 
$-R$. One then considers the limiting process where the curvature
of the hyperbolic space tends to zero and the limiting conformal 
structure near the point $p$ is a Gibbons-Hawking metric. 
From the above, it is clear that an $S^1$ invariant instanton 
of weight $s$ on $R^4$ defines a monopole on $H^3(R)$. Because 
there is a scale change in the circle, the Higgs field tends to 
$R^{-1} s$ at infinity. Then taking $R = s$, the monopoles on 
$H^3$ are reinterpreted as a monopole on $H^3(s^{-1})$ \cite
{Atiyah}. As 
$s \rightarrow \infty $, the hyperbolic space $H^3(s^{-1})$ 
tends to the flat space $R^3$ and in this limit, the hyperbolic 
monopole becomes indeed an ordinary monopole. 

In this note, we have discussed various different ways of 
obtaining 
Eguchi-Hanson metrics with or without cosmological constant 
term through
some limiting procedure. We also discussed the generalized 
Gibbons-Hawking ansatz of Lebrun in the context of non linear 
partial differential equation namely the continual Toda equation
and its linearization arising for rotational killing symmetry.
For the example discussed in the text, in the context 
of rotational killing symmetry, one obtains the hyperbolic 
analog of Gibbons-Hawking metric and again one reproduces 
ALE metric as a limit of the self-dual metrics on the connected
sum of $n CP_2$. This might be interesting in the context of 
$D$-branes on resoved ALE spaces \cite{Moore}.  


{\bf Acknowledgments}

We gratefully acknowledge the Abdus Salam International Centre
for Theoretical Physics for an Associateship, under which a part  
of this work has been done.

\medskip



\end{document}